%% file: main.tex
\documentclass[12pt,a4paper]{article}

\usepackage{ifthen} 
\newboolean{pdflatex}
\setboolean{pdflatex}{false} 

\newboolean{articletitles}
\setboolean{articletitles}{true} 

\newboolean{uprightparticles}
\setboolean{uprightparticles}{false} 

\newboolean{inbibliography}
\setboolean{inbibliography}{false} 

\newboolean{prlstyle}
\setboolean{prlstyle}{false}

\newboolean{wordcount}
\setboolean{wordcount}{false} 

\input{preamble}
\usepackage{longtable} 
\usepackage{multirow}
\usepackage{amsmath}
\usepackage[makeroom]{cancel}

\begin{document}

\renewcommand{\thefootnote}{\fnsymbol{footnote}}
\setcounter{footnote}{1}

\input{title-LHCb-PAPER}

\renewcommand{\thefootnote}{\arabic{footnote}}
\setcounter{footnote}{0}


\pagestyle{plain} 
\setcounter{page}{1}
\pagenumbering{arabic}

\input{introduction}

\input{detector}

\input{selection}

\input{results}

\input{conclusions}

\input{acknowledgements}

\addcontentsline{toc}{section}{References}
\setboolean{inbibliography}{true}
\bibliographystyle{LHCb}
\bibliography{main,LHCb-PAPER,LHCb-CONF,LHCb-DP,LHCb-TDR}

\newpage
\input{LHCb_Authorship_flat_06-Jul-2016.tex}

\end{document}

%% file: preamble.tex
\usepackage[top=1in, bottom=1.25in, left=1in, right=1in]{geometry}

\usepackage{microtype}
\usepackage{lineno}  
\usepackage{xspace} 
\usepackage{caption} 

\usepackage{graphicx}  
\usepackage{color}
\usepackage{colortbl}
\graphicspath{{./figs/}} 

\usepackage{amsmath} 
\usepackage{amssymb}
\usepackage{amsfonts}
\usepackage{upgreek} 

\newcommand*\patchAmsMathEnvironmentForLineno[1]{%
\expandafter\let\csname old#1\expandafter\endcsname\csname #1\endcsname
\expandafter\let\csname oldend#1\expandafter\endcsname\csname
end#1\endcsname
 \renewenvironment{#1}%
   {\linenomath\csname old#1\endcsname}%
   {\csname oldend#1\endcsname\endlinenomath}%
}
\newcommand*\patchBothAmsMathEnvironmentsForLineno[1]{%
  \patchAmsMathEnvironmentForLineno{#1}%
  \patchAmsMathEnvironmentForLineno{#1*}%
}
\AtBeginDocument{%
\patchBothAmsMathEnvironmentsForLineno{equation}%
\patchBothAmsMathEnvironmentsForLineno{align}%
\patchBothAmsMathEnvironmentsForLineno{flalign}%
\patchBothAmsMathEnvironmentsForLineno{alignat}%
\patchBothAmsMathEnvironmentsForLineno{gather}%
\patchBothAmsMathEnvironmentsForLineno{multline}%
\patchBothAmsMathEnvironmentsForLineno{eqnarray}%
}

\usepackage{hyperref}    
\usepackage[all]{hypcap} 

\input{lhcb-symbols-def} 

\usepackage{cite} 
\usepackage{mciteplus}

%% file: lhcb-symbols-def.tex
\def\lhcb {\mbox{LHCb}\xspace}

\def\MagUp {\mbox{\em Mag\kern -0.05em Up}\xspace}

\ifthenelse{\boolean{uprightparticles}}%
{

 \def\Ppi         {\ensuremath{\uppi}\xspace}

 \def\PDelta      {\ensuremath{\Delta}\xspace}                 
 \def\PXi      {\ensuremath{\Xi}\xspace}                 
 \def\PLambda      {\ensuremath{\Lambda}\xspace}                 
 \def\PSigma      {\ensuremath{\Sigma}\xspace}                 
 \def\POmega      {\ensuremath{\Omega}\xspace}                 
 \def\PUpsilon      {\ensuremath{\Upsilon}\xspace}                 
 
 \def\PB      {\ensuremath{\mathrm{B}}\xspace}                 
                  
 \def\PD      {\ensuremath{\mathrm{D}}\xspace}

 \def\PK      {\ensuremath{\mathrm{K}}\xspace}

 \def\PV      {\ensuremath{\mathrm{V}}\xspace}

 \def\Pb      {\ensuremath{\mathrm{b}}\xspace}                 
 \def\Pc      {\ensuremath{\mathrm{c}}\xspace}                 
 \def\Pd      {\ensuremath{\mathrm{d}}\xspace}

 \def\Pi      {\ensuremath{\mathrm{i}}\xspace}

 \def\Pp      {\ensuremath{\mathrm{p}}\xspace}

 \def\Ps      {\ensuremath{\mathrm{s}}\xspace}                 
                  
 \def\Pu      {\ensuremath{\mathrm{u}}\xspace}

}
{

 \def\Ppi         {\ensuremath{\pi}\xspace}

 \mathchardef\PDelta="7101
 \mathchardef\PXi="7104
 \mathchardef\PLambda="7103
 \mathchardef\PSigma="7106
 \mathchardef\POmega="710A
 \mathchardef\PUpsilon="7107
                  
 \def\PB      {\ensuremath{B}\xspace}                 
                  
 \def\PD      {\ensuremath{D}\xspace}

 \def\PK      {\ensuremath{K}\xspace}

 \def\PV      {\ensuremath{V}\xspace}

 \def\Pb      {\ensuremath{b}\xspace}                 
 \def\Pc      {\ensuremath{c}\xspace}                 
 \def\Pd      {\ensuremath{d}\xspace}

 \def\Pi      {\ensuremath{i}\xspace}

 \def\Pp      {\ensuremath{p}\xspace}

 \def\Ps      {\ensuremath{s}\xspace}                 
                  
 \def\Pu      {\ensuremath{u}\xspace}

}

\makeatletter
\ifcase \@ptsize \relax
  \newcommand{\miniscule}{\@setfontsize\miniscule{4}{5}}
\or
  \newcommand{\miniscule}{\@setfontsize\miniscule{5}{6}}
\or
  \newcommand{\miniscule}{\@setfontsize\miniscule{5}{6}}
\fi
\makeatother

\DeclareRobustCommand{\optbar}[1]{\shortstack{{\miniscule (\rule[.5ex]{1.25em}{.18mm})}
  \\ [-.7ex] $#1$}}

\def\uquark    {{\ensuremath{\Pu}}\xspace}
\def\uquarkbar {{\ensuremath{\overline \uquark}}\xspace}

\def\dquark    {{\ensuremath{\Pd}}\xspace}

\def\squark    {{\ensuremath{\Ps}}\xspace}

\def\cquark    {{\ensuremath{\Pc}}\xspace}

\def\bquark    {{\ensuremath{\Pb}}\xspace}

\def\pion   {{\ensuremath{\Ppi}}\xspace}

\def\pip    {{\ensuremath{\pion^+}}\xspace}
\def\pim    {{\ensuremath{\pion^-}}\xspace}

\def\kaon    {{\ensuremath{\PK}}\xspace}
  \def\Kbar    {{\kern 0.2em\overline{\kern -0.2em \PK}{}}\xspace}

\def\KorKbar    {\kern 0.18em\optbar{\kern -0.18em K}{}\xspace}

\def\Kp      {{\ensuremath{\kaon^+}}\xspace}
\def\Km      {{\ensuremath{\kaon^-}}\xspace}

  \def\Dbar    {{\kern 0.2em\overline{\kern -0.2em \PD}{}}\xspace}

\def\DorDbar    {\kern 0.18em\optbar{\kern -0.18em D}{}\xspace}

\def\B       {{\ensuremath{\PB}}\xspace}
\def\Bbar    {{\ensuremath{\kern 0.18em\overline{\kern -0.18em \PB}{}}}\xspace}

\def\BorBbar    {\kern 0.18em\optbar{\kern -0.18em B}{}\xspace}
\def\Bz      {{\ensuremath{\B^0}}\xspace}

\def\Bu      {{\ensuremath{\B^+}}\xspace}

\def\Bp      {{\ensuremath{\Bu}}\xspace}

\def\Bd      {{\ensuremath{\B^0}}\xspace}
\def\Bs      {{\ensuremath{\B^0_\squark}}\xspace}

  \def\Y#1S{\ensuremath{\PUpsilon{(#1S)}}\xspace}

\def\proton      {{\ensuremath{\Pp}}\xspace}
\def\antiproton  {{\ensuremath{\overline \proton}}\xspace}

\def\Deltares    {{\ensuremath{\PDelta}}\xspace}

\def\Lz          {{\ensuremath{\PLambda}}\xspace}
\def\Lbar        {{\ensuremath{\kern 0.1em\overline{\kern -0.1em\PLambda}}}\xspace}
\def\LorLbar    {\kern 0.18em\optbar{\kern -0.18em \PLambda}{}\xspace}

\def\Lb      {{\ensuremath{\Lz^0_\bquark}}\xspace}
\def\Lbbar   {{\ensuremath{\Lbar{}^0_\bquark}}\xspace}
\def\Lc      {{\ensuremath{\Lz^+_\cquark}}\xspace}

\newcommand{\decay}[2]{\ensuremath{#1\!\to #2}\xspace}         

\def\to                 {\ensuremath{\rightarrow}\xspace}

\def\CP                {{\ensuremath{C\!P}}\xspace}

\def\CPV               {{\mbox{CPV}}\xspace}
\def\PV                 {{\mbox{PV}}\xspace}

\newcommand{\AT}{{\ensuremath{A_{\widehat{T}}}}\xspace}
\newcommand{\ATbar}{{\ensuremath{\kern 0.1em\overline{\kern -0.1em A}_{\widehat{T}}}}\xspace}
\newcommand{\aCPTodd}{{\ensuremath{a_\CP^{\widehat{T}\text{-odd}}}}\xspace}
\newcommand{\aPTodd}{{\ensuremath{a_P^{\widehat{T}\text{-odd}}}}\xspace}

\newcommand{\CT}{{\ensuremath{C_{\widehat{T}}}}\xspace}
\newcommand{\CTbar}{{\ensuremath{\kern 0.1em\overline{\kern -0.1em C}_{\widehat{T}}}}\xspace}

\def\LbTopKpipi   {\decay{\Lb}{\proton\Km\pip\pim}}

\def\LbToppiKK    {\decay{\Lb}{\proton\pim\Kp\Km}}
\def\LbToppipipi  {\decay{\Lb}{\proton\pim\pip\pim}}

\def\LbToLcpiCS   {\decay{\Lb}{\Lc(\decay{}{\proton\Km\pip})\pim}}
\def\LbToLcpi    {\decay{\Lb}{\Lc\pim}}

\def\T          {\ensuremath{T}\xspace}
\def\That       {\ensuremath{\widehat{T}}\xspace}
\def\atv    {\ensuremath{a^{\widehat{T}-\text{odd}}_{\CP}}\xspace}

\def\C#1      {\ensuremath{\mathcal{C}_{#1}}\xspace}                       
\def\Cp#1     {\ensuremath{\mathcal{C}_{#1}^{'}}\xspace}                    
\def\Ceff#1   {\ensuremath{\mathcal{C}_{#1}^{\mathrm{(eff)}}}\xspace}        
\def\Cpeff#1  {\ensuremath{\mathcal{C}_{#1}^{'\mathrm{(eff)}}}\xspace}       
\def\Ope#1    {\ensuremath{\mathcal{O}_{#1}}\xspace}                       
\def\Opep#1   {\ensuremath{\mathcal{O}_{#1}^{'}}\xspace}                    

\newcommand{\tev}{\ifthenelse{\boolean{inbibliography}}{\ensuremath{~T\kern -0.05em eV}\xspace}{\ensuremath{\mathrm{\,Te\kern -0.1em V}}}\xspace}
\newcommand{\gev}{\ensuremath{\mathrm{\,Ge\kern -0.1em V}}\xspace}
\newcommand{\mev}{\ensuremath{\mathrm{\,Me\kern -0.1em V}}\xspace}
\newcommand{\kev}{\ensuremath{\mathrm{\,ke\kern -0.1em V}}\xspace}
\newcommand{\ev}{\ensuremath{\mathrm{\,e\kern -0.1em V}}\xspace}
\newcommand{\gevc}{\ensuremath{{\mathrm{\,Ge\kern -0.1em V\!/}c}}\xspace}
\newcommand{\mevc}{\ensuremath{{\mathrm{\,Me\kern -0.1em V\!/}c}}\xspace}
\newcommand{\gevcc}{\ensuremath{{\mathrm{\,Ge\kern -0.1em V\!/}c^2}}\xspace}
\newcommand{\gevgevcccc}{\ensuremath{{\mathrm{\,Ge\kern -0.1em V^2\!/}c^4}}\xspace}
\newcommand{\mevcc}{\ensuremath{{\mathrm{\,Me\kern -0.1em V\!/}c^2}}\xspace}

\def\invfb   {\ensuremath{\mbox{\,fb}^{-1}}\xspace}

\def\ps   {\ensuremath{{\rm \,ps}}\xspace}

\newcommand{\chisq}{\ensuremath{\chi^2}\xspace}

\def\gsim{{~\raise.15em\hbox{$>$}\kern-.85em
          \lower.35em\hbox{$\sim$}~}\xspace}
\def\lsim{{~\raise.15em\hbox{$<$}\kern-.85em
          \lower.35em\hbox{$\sim$}~}\xspace}

\def\mrad{\ensuremath{\rm \,mrad}\xspace}

\def\tell1  {TELL1\xspace}
\def\ukl1   {UKL1\xspace}

\newcommand{\eg}{\mbox{\itshape e.g.}\xspace}
\newcommand{\ie}{\mbox{\itshape i.e.}\xspace}


%% file: title-LHCb-PAPER.tex
\begin{titlepage}
\pagenumbering{roman}

\vspace*{-1.5cm}
\centerline{\large EUROPEAN ORGANIZATION FOR NUCLEAR RESEARCH (CERN)}
\vspace*{1.5cm}
\noindent
\begin{tabular*}{\linewidth}{lc@{\extracolsep{\fill}}r@{\extracolsep{0pt}}}
\ifthenelse{\boolean{pdflatex}}
{\vspace*{-2.7cm}\mbox{\!\!\!\includegraphics[width=.14\textwidth]{lhcb-logo.pdf}} & &}%
{\vspace*{-1.2cm}\mbox{\!\!\!\includegraphics[width=.12\textwidth]{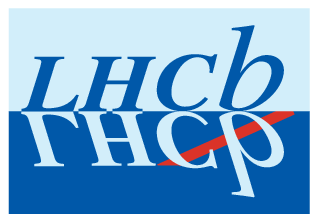}} & &}%
\\
 & & CERN-EP-2016-212 \\  
 & & LHCb-PAPER-2016-030 \\  
 & & October 27, 2017 \\ 
\end{tabular*}

\vspace*{2.0cm}

{\normalfont\bfseries\boldmath\huge
\begin{center}
  Measurement of matter-antimatter differences in beauty baryon decays
\end{center}
}

\vspace*{1.0cm}

\begin{center}
The LHCb collaboration\footnote{Authors are listed at the end of this letter.}
\end{center}

\vspace{\fill}

\begin{abstract}
  \noindent
  Differences in the behaviour of matter and antimatter have been observed in $K$ and $B$ meson decays, but not yet in any baryon decay.
  Such differences are associated with the non-invariance of fundamental interactions under the combined
  charge-conjugation and parity transformations, known as \CP violation.
  Using data from the \lhcb experiment at the Large Hadron Collider, a search is made for \CP-violating asymmetries
  in the decay angle distributions of \Lb baryons decaying to $\proton\pim\pip\pim$ and $\proton\pim\Kp\Km$ final states.
  These four-body hadronic decays
  are a promising place to search for sources of \CP violation both within and beyond the Standard Model of particle physics.
  We find evidence for \CP violation in \Lb to $\proton\pim\pip\pim$ decays with a statistical significance
  corresponding to 3.3 standard deviations including systematic uncertainties.
  This represents the first evidence for \CP violation in the baryon sector.
\end{abstract}

\vspace*{2.0cm}

\begin{center}
  Published in Nature Physics 13, 391--396 (2017)
\end{center}

\vspace{\fill}

{\footnotesize
\centerline{\copyright~CERN on behalf of the \lhcb collaboration, licence \href{http://creativecommons.org/licenses/by/4.0/}{CC-BY-4.0}.}}
\vspace*{2mm}

\end{titlepage}


\newpage
\setcounter{page}{2}
\mbox{~}

\cleardoublepage

%% file: introduction.tex
The asymmetry between matter and antimatter is related to the violation of the \CP symmetry (\CPV), where $C$ and $P$ are the charge-conjugation and parity operators.
\CP violation is accommodated in the Standard Model (SM) of particle physics by the Cabibbo-Kobayashi-Maskawa (CKM)
mechanism that describes the transitions between up- and down-type quarks~\cite{PhysRevLett.10.531,Kobayashi:1973fv}, in which quark decays proceed by the emission of a virtual $W$ boson and where the phases of the couplings change sign between quarks and antiquarks.
However, the amount of \CPV predicted by the CKM mechanism is not sufficient to explain our matter-dominated Universe~\cite{Sakharov:1967dj,Riotto:1998bt} and other sources of \CPV are expected to exist.
The initial discovery of \CPV was in neutral $K$ meson decays~\cite{PhysRevLett.13.138}, and more recently it has been observed in \Bd~\cite{Aubert:2001nu,Abe:2001xe},
\Bp~\cite{Aubert:2008bd,Poluektov:2010wz,Aaij:2012kz,Aaij:2013sfa}, and \Bs~\cite{Aaij:2013iua} meson decays,
but it has never been observed in the decays of any baryon.
Decays of the \Lb ($bud$) baryon to final states consisting of hadrons with no charm quarks are predicted to have non-negligible \CP asymmetries in the SM, as large as 20\% for certain three-body decay modes~\cite{Hsiao:2014mua}.
It is important to measure the size and nature of these \CP asymmetries in as many decay modes as possible, to determine whether they are consistent with the CKM mechanism or, if not, what extensions to the SM would be required to explain
them~\cite{Bensalem:2000hq,Bensalem:2002ys,Bensalem:2002pz}.

\begin{figure}[!h]
\centering
\includegraphics[width=0.9\textwidth]{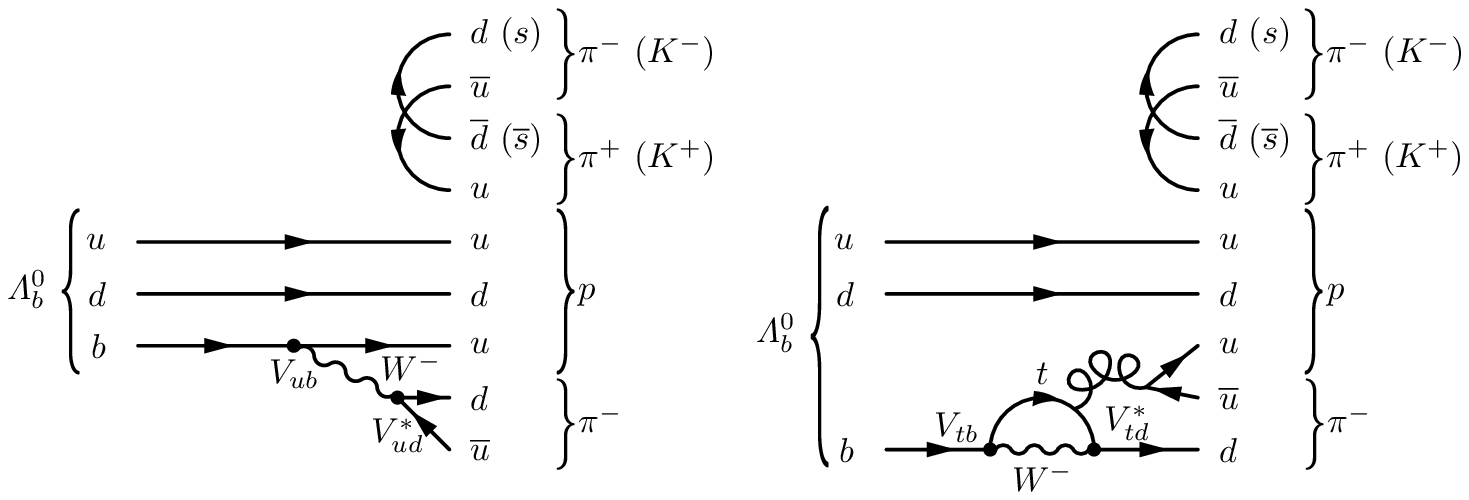}
\caption{\label{fig:feynman}
{\bf Dominant Feynman diagrams for \boldmath{\LbToppipipi} and \boldmath{\LbToppiKK} transitions.}
The two diagrams show the transitions that contribute most strongly to \LbToppipipi and \LbToppiKK decays.
In both cases, a pair of $\pip\pim$ ($\Kp\Km$) is produced by gluon emission from the light quarks (\uquark,\dquark).
The difference is in the \bquark quark decay that happens on the left through a virtual $W^-$ boson emission (``tree diagram'') and on the right as a virtual $W^-$ boson emission and absorption together with a gluon emission (``loop diagram'').
The magnitudes of the two amplitudes are expected to be comparable, and each is proportional to the product of the CKM matrix elements involved, which are shown in the figure.}
\end{figure}

The decay processes studied in this article, \LbToppipipi and \LbToppiKK, are mediated by the weak interaction and governed mainly by two amplitudes, expected to be of similar magnitude, from different diagrams describing quark-level $\bquark \to \uquark \uquarkbar \dquark$ transitions, as shown in Fig.~\ref{fig:feynman}.
Throughout this paper the inclusion of charge-conjugate reactions is implied, unless otherwise indicated.
\CPV could arise from the interference of two amplitudes with relative phases that differ between particle and antiparticle decays, leading to differences in the \Lb and \Lbbar decay rates.
The main source of this effect in the SM would be the large relative phase (referred to as $\alpha$ in the literature) between the product of the CKM matrix elements $V_{ub}V_{ud}^*$ and $V_{tb}V_{td}^*$, which are present in the different diagrams depicted in Fig.~\ref{fig:feynman}.
Parity violation (\PV) is also expected in weak interactions, but has never been observed in \Lb decays.

To search for \CP-violating effects one needs to measure \CP-odd observables, which can be done by studying asymmetries in the \That operator.
This is a unitary operator that reverses both the momentum and spin three-vectors~\cite{Sachs,Branco:1999fs}, and is different from the antiunitary time-reversal operator \T~\cite{Durieux:2015zwa,Durieux:2016nqr} that also exchanges initial and
final states.
A non-zero \CP-odd observable implies \CP violation and similar considerations apply to $P$-odd observables and parity violation~\cite{gasiorowicz1966elementary}.
Furthermore, different values of $P$-odd observables for a decay and its charge conjugate would imply \CPV.
In this paper, scalar triple products of final-state particle momenta in the \Lb centre-of-mass frame are studied to search for $P$- and \CP-violating effects in four-body decays.
These are defined as $\CT=\vec{p}_\proton\cdot(\vec{p}_{h_1^-}\times\vec{p}_{h_2^+})$ for \Lb and
 $\CTbar=\vec{p}_\antiproton\cdot(\vec{p}_{h_1^+}\times\vec{p}_{h_2^-})$ for \Lbbar,
where $h_1$ and $h_2$ are final state hadrons: $h_1=\pi$ and $h_2=K$ for \LbToppiKK and $h_1=h_2=\pi$
for \LbToppipipi. In the latter case there is an inherent ambiguity in the choice of the pion for $h_1$ that is resolved
by taking that with the larger momentum in the \Lb rest frame, referred to as $\pi_\textrm{fast}$.
The following asymmetries may then be defined~\cite{Bigi:2001sg,Gronau:2011cf}:
\begin{align}
\label{eq:AT}
\AT(\CT) &= \frac{N(\CT>0)-N(\CT<0)}{N(\CT>0)+N(\CT<0)},\\
\label{eq:ATb}
\ATbar(\CTbar) &= \frac{\overline{N}(-\CTbar>0)-\overline{N}(-\CTbar<0)}{\overline{N}(-\CTbar>0)+\overline{N}(-\CTbar<0)},
\end{align}
where $N$ and $\overline{N}$ are the numbers of \Lb and \Lbbar decays.
These asymmetries are $P$-odd and \That-odd and so change sign under $P$ or \That transformations, \ie $\AT (\CT) = - \AT (-\CT)$ or $\ATbar (\CTbar) = - \ATbar (-\CTbar)$.
The $P$- and \CP-violating observables are defined as
\begin{align}
\label{eq:ATV}
\aPTodd = \frac{1}{2}\left(\AT +\ATbar\right),\hspace{3cm}
\aCPTodd = \frac{1}{2}\left(\AT- \ATbar\right),
\end{align}
and a significant deviation from zero would signal \PV or \CPV, respectively.

Searches for \CPV with triple-product asymmetries are particularly suited to \Lb four-body decays to hadrons with no charm quark~\cite{Gronau:2015gha} thanks to the rich resonant substructure, dominated by $\Deltares(1232)^{++}\to\proton\pip$ and $\rho(770)^{0}\to\pip\pim$ resonances in the \LbToppipipi final state.
The observable \aCPTodd is sensitive to the interference of \That-even and \That-odd amplitudes with different \CP-odd (``weak'') phases.
Unlike the overall asymmetry in the decay rate that is sensitive to the interference of \That-even amplitudes,
\aCPTodd does not require a non-vanishing difference in the \CP-invariant (``strong'') phase between the contributing amplitudes~\cite{PhysRevD.39.3339,Durieux:2015zwa}.
The observables \AT, \ATbar, \aPTodd and \aCPTodd are, by construction, largely insensitive to particle-antiparticle
production asymmetries and detector-induced charge asymmetries~\cite{Aaij:2014qwa}.

This article describes measurements of the \CP- and $P$-violating asymmetries introduced in equation~\eqref{eq:ATV} in
\LbToppipipi and \LbToppiKK decays.
The asymmetries are measured first for the entire phase space of the decay, integrating over all possible final state
configurations, and then in different regions of phase space so as to enhance sensitivity to localised \CPV.
The analysis is performed using proton-proton collision data collected by the \lhcb detector, corresponding to $3.0\invfb$
of integrated luminosity at centre-of-mass energies of 7 and 8\tev, and exploits the copious production of \Lb baryons at
the LHC, which constitutes around 20\% of all \bquark hadrons produced~\cite{Aaij:2011jp}.
Control samples of \LbTopKpipi and \LbToLcpi decays,
with $\Lc$ decaying to $\proton\Km\pip$, $\proton\pim\pip$, and $\proton\Km\Kp$ final states,
are used to optimise the event selection and study systematic effects;
the most abundant control sample consists of $\Lb\to\Lc(\to\proton\Km\pip)\pim$ decays mediated by \decay{\bquark}{\cquark}
quark transitions in which no \CPV is expected~\cite{LHCb-PAPER-2014-004}.
To avoid introducing biases in the results, all aspects of the analysis, including the selection, phase space regions,
and procedure used to determine the statistical significance of the results, were fixed before the data were examined.

%% file: detector.tex
The \lhcb detector~\cite{Alves:2008zz,LHCb-DP-2014-002} is designed to collect data of \bquark-hadron decays produced from proton-proton collisions at the Large Hadron Collider.
It instruments a region around the proton beam axis, covering the polar angles between 10 and 250\mrad,
where approximately 24\% of the \bquark-hadron decays occur~\cite{LHCB:1995aa}.
The detector includes a high-precision tracking system with a dipole magnet, providing measurements of the momentum and decay vertex position of particle decays.
Different types of charged particles are distinguished using information from two ring-imaging Cherenkov detectors, a calorimeter and a muon system.
Simulated samples of \Lb signal modes and control samples are used in this analysis to verify the experimental method and to study certain systematic effects.
These simulated events model the experimental conditions in detail, including the proton-proton collision, the decays of the particles, and the response of the detector.
The software used is described in Refs.~\cite{Sjostrand:2006za,*Sjostrand:2007gs,LHCb-PROC-2010-056,Lange:2001uf,Allison:2006ve, *Agostinelli:2002hh,LHCb-PROC-2011-006}.
The online event selection is performed by a trigger system that takes fast decisions about which events to record.
It consists of a hardware stage, based on information from the calorimeter and muon systems, followed by a software stage, which applies a full event reconstruction.
The software trigger requires \Lb candidates to be consistent with a \bquark-hadron decay topology, with tracks originating from a secondary vertex detached from the primary $pp$ collision point.
The mean \Lb lifetime is 1.5\ps~\cite{PDG2014}, which corresponds to a typical flight distance of a few millimetres in \lhcb.

%% file: selection.tex
The $\Lb\to p\pim h^+h^-$  candidates are formed by combining tracks identified as protons, pions, or kaons that originate from a common vertex.
The proton or antiproton identifies the candidate as a \Lb or \Lbbar.
There are backgrounds from \bquark-hadron decays to charm hadrons that are suppressed by reconstructing the appropriate two- or three-body invariant masses, and requiring them to differ from the known charm hadron masses by at least three times the experimental resolution.
For the \LbToLcpi control mode, only the $\decay{\Lb}{\proton h^+h^-\pim}$ events with reconstructed
$\proton h^+h^-$ invariant mass between $2.23$ and $2.31$ \gevcc are retained.

A boosted decision tree (BDT) classifier~\cite{Breiman} is constructed from a set of kinematic variables
 that discriminate between signal and background.
The signal and background training samples used for the BDT are derived from the \LbTopKpipi control sample, since its kinematics and topology are similar to the decays under study;
background in this sample is subtracted with the $sPlot$ technique~\cite{Pivk:2004ty}, a statistical technique to disentangle signal and background contributions.
The background training sample consists of candidates that lie far from the signal mass peak, between 5.85 and 6.40 \gevcc.
The control modes $\Lb\to\Lc(\to p\pip\pim)\pim$ and $\Lb\to\Lc(\to p\Km\Kp)\pim$ are used to optimise the particle identification criteria for the signal mode with the same final state.
For events in which multiple candidates pass all selection criteria for a given mode, one candidate is retained at random and the rest discarded.

\begin{figure}[tb]
\centering
\small
\includegraphics[width=0.48\linewidth]{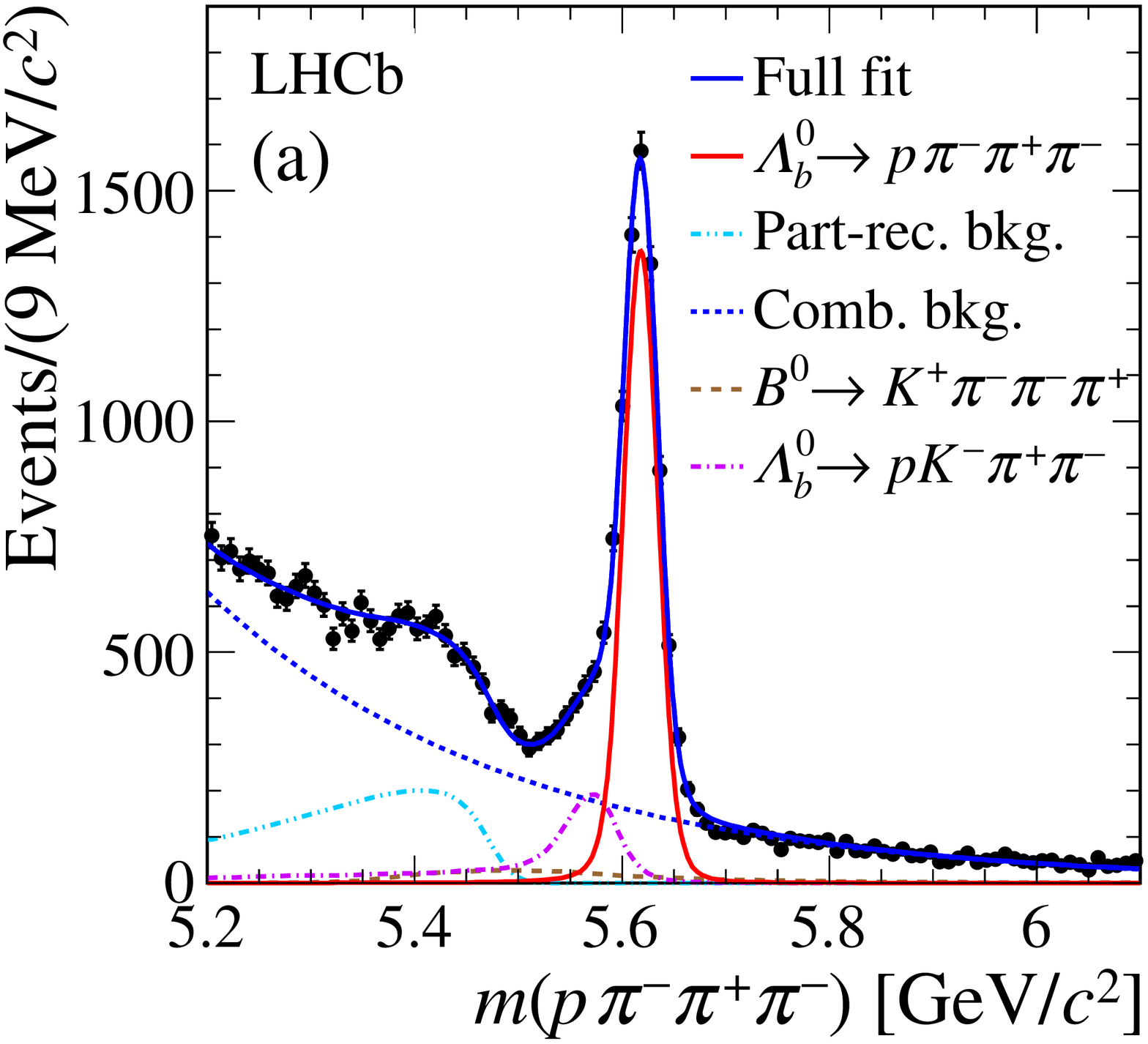}
\includegraphics[width=0.48\linewidth]{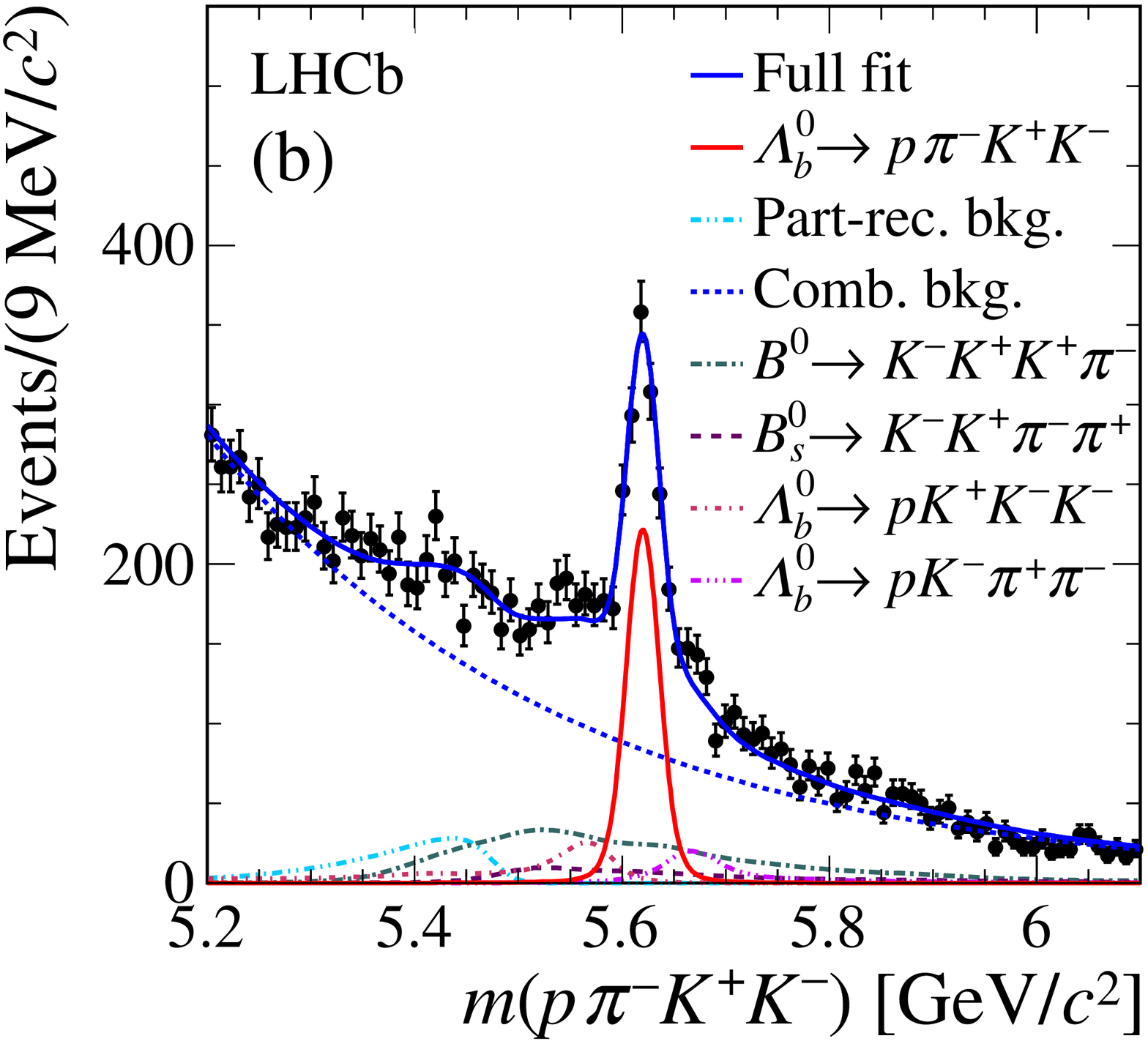}
\caption{\label{fig:p3h_massfit} {\bf Reconstructed invariant mass fits used to extract the signal yields.} The invariant mass distributions for (a) \LbToppipipi  and
  (b) \LbToppiKK decays are shown. A fit is overlaid on top of the data points, with solid and dotted lines describing the projections of the fit results for each of the components described in the text and listed in the legend.
  Uncertainties on the data points are statistical only and are calculated assuming Poisson distributed entries.}
\end{figure}

Unbinned extended maximum likelihood fits to the $\proton\pim\pip\pim$ and the $\proton\pim\Kp\Km$ invariant mass distributions are shown in Fig.~\ref{fig:p3h_massfit}.
The invariant mass distribution of the \Lb signal is modelled by a Gaussian core with power law tails~\cite{Skwarnicki:1986xj} with the mean and the width of the Gaussian determined from the fit to data.
The combinatorial background is modelled by an exponential distribution with the rate parameter extracted from data.
All other parameters of the fit model are taken from simulations except the yields.
Partially reconstructed \Lb decays are described by an empirical function~\cite{Albrecht:1990am} convolved with a Gaussian function to account for resolution effects.
The shapes of backgrounds from other \bquark-hadron decays due to incorrectly identified particles,
\eg kaons identified as pions or protons identified as kaons, are modelled using simulated events.
These consist mainly of \LbTopKpipi and $\Bz\to\Kp\pim\pim\pip$ decays for the \LbToppipipi sample and of
similar final states for the \LbToppiKK sample, as shown in Fig.~\ref{fig:p3h_massfit}.
The yields of these contributions are obtained from fits to
data reconstructed under the appropriate mass hypotheses for the final-state particles.
The signal yields of \LbToppipipi and \LbToppiKK are $6646\pm105$ and $1030\pm56$, respectively.
This is the first observation of these decay modes.

%% file: results.tex
Signal candidates are split into four categories according to \Lb or \Lbbar flavour and the sign of \CT or $\CTbar$ in order to calculate the asymmetries defined in equations~(\ref{eq:AT}) and~(\ref{eq:ATb}).
The reconstruction efficiency for signal candidates with $\CT>0$ is identical to that with $\CT<0$ within the statistical uncertainties of the control sample, and likewise for \CTbar,
which indicates that the detector and the reconstruction program do not bias this measurement.
This check is performed both on the \decay{\Lb}{\Lc(\proton\Km\pip)\pim} data control sample and on large samples of simulated events, using yields about 30 times those found in data, which are generated with no \CP asymmetry.
The \CP asymmetry measured in the control sample is $\aCPTodd(\Lc\pim) = (0.15\pm0.31)\%$, compatible
with \CP symmetry.
The asymmetries \AT and \ATbar in the signal samples are measured with a simultaneous unbinned maximum likelihood fit to the invariant mass distributions
of the different signal categories, and are found to be uncorrelated.
Corresponding asymmetries for each of the background components are also measured in the fit;
they are found to be consistent with zero, and do not lead to significant systematic uncertainties in the signal asymmetries. The values of \aCPTodd and \aPTodd are then calculated from \AT and \ATbar.

\begin{figure}[t]
\centering
\small
\includegraphics[width=0.7\textwidth]{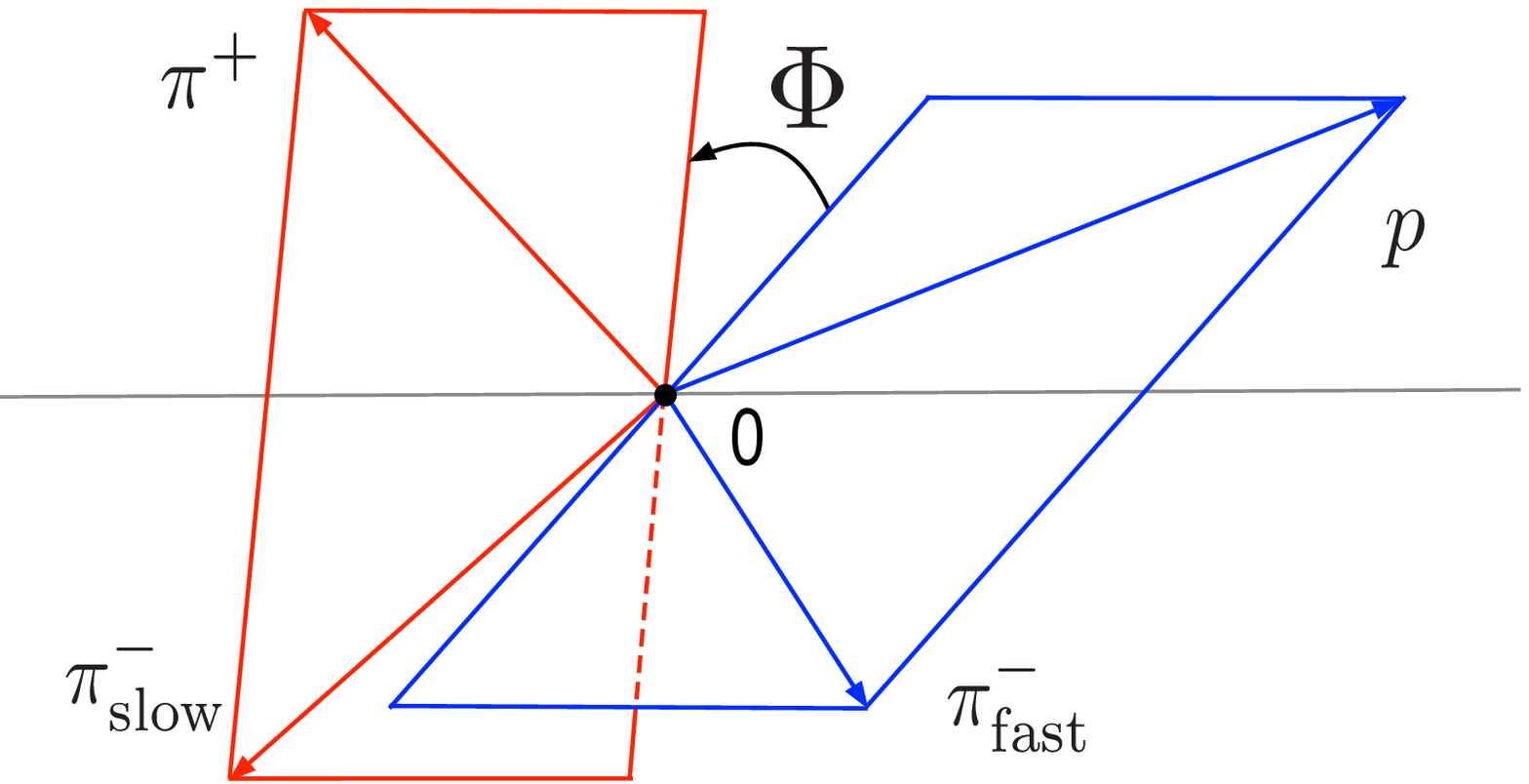}
\caption{\small\label{fig:decayPlane} {\bf Definition of the $\Phi$ angle.} The decay planes formed by the
$\proton\pi^-_\textrm{fast}$ and the $\pi^-_\textrm{slow}\pip$ systems in the \Lb rest frame. The momenta
 of the particles, represented by vectors, determine the two decay planes and the angle $\Phi \in [-\pi,\pi]$~\cite{Durieux:2015zwa} measures their relative orientation.}
\end{figure}

In four-body particle decays, the \CP asymmetries may vary over the phase space due to resonant contributions or their interference effects, possibly cancelling when integrated over the whole phase space.
Therefore, the asymmetries are measured in different regions of phase space for the \LbToppipipi decay using two binning schemes, defined before examing the data.
Scheme A, defined in Table~\ref{tab:definitionscheme_Lb2p3pi}, is designed to isolate regions of phase space according to their dominant resonant contributions.
Scheme B exploits in more detail the interference of contributions which could be visible as a function of the angle
$\Phi$ between the decay planes formed by the $\proton\pi^-_\textrm{fast}$ and the $\pi^-_\textrm{slow}\pip$ systems, as illustrated in Fig.~\ref{fig:decayPlane}.
Scheme B has 10 non-overlapping bins of width $\pi/10$ in $|\Phi|$.
For every bin in each of the schemes, the \Lb efficiencies for $\CT>0$ and $\CT<0$ are compared and found to be equal within uncertainties, and likewise the \Lbbar efficiencies for $\CTbar>0$ and $\CTbar<0$.
The analysis technique is validated on the \decay{\Lb}{\Lc(\proton\Km\pip)\pim} control sample, for which the angle $\Phi$ is defined by the decay planes of the $\proton\Km$ and $\pip\pim$ pairs, and on simulated signal events.

\begin{table}[b]
\small
\centering
\caption{\small\label{tab:definitionscheme_Lb2p3pi}{\bf Definition of binning scheme A for the decay mode \boldmath{\LbToppipipi}.}
Binning scheme A is defined to exploit interference patterns arising from the resonant structure of the decay.
Bins 1-4 focus on the region dominated by the $\Deltares(1232)^{++}\to\proton\pip$ resonance.
The other eight bins are defined to study regions where $\proton\pim$ resonances are present (5--8)
on either side of the $\rho(770)^{0}\to\pip\pim$ resonances (5--12).
Further splitting for $|\Phi|$ lower or greater than $\pi/2$ is done to reduce potential dilution of asymmetries, as suggested in Ref.~\cite{Durieux:2015zwa}.
Masses are in units of \gevcc.}
\input binning_definition_tab.tex
\end{table}

\begin{figure}[tb]
\centering
\small
\includegraphics[width=0.48\textwidth]{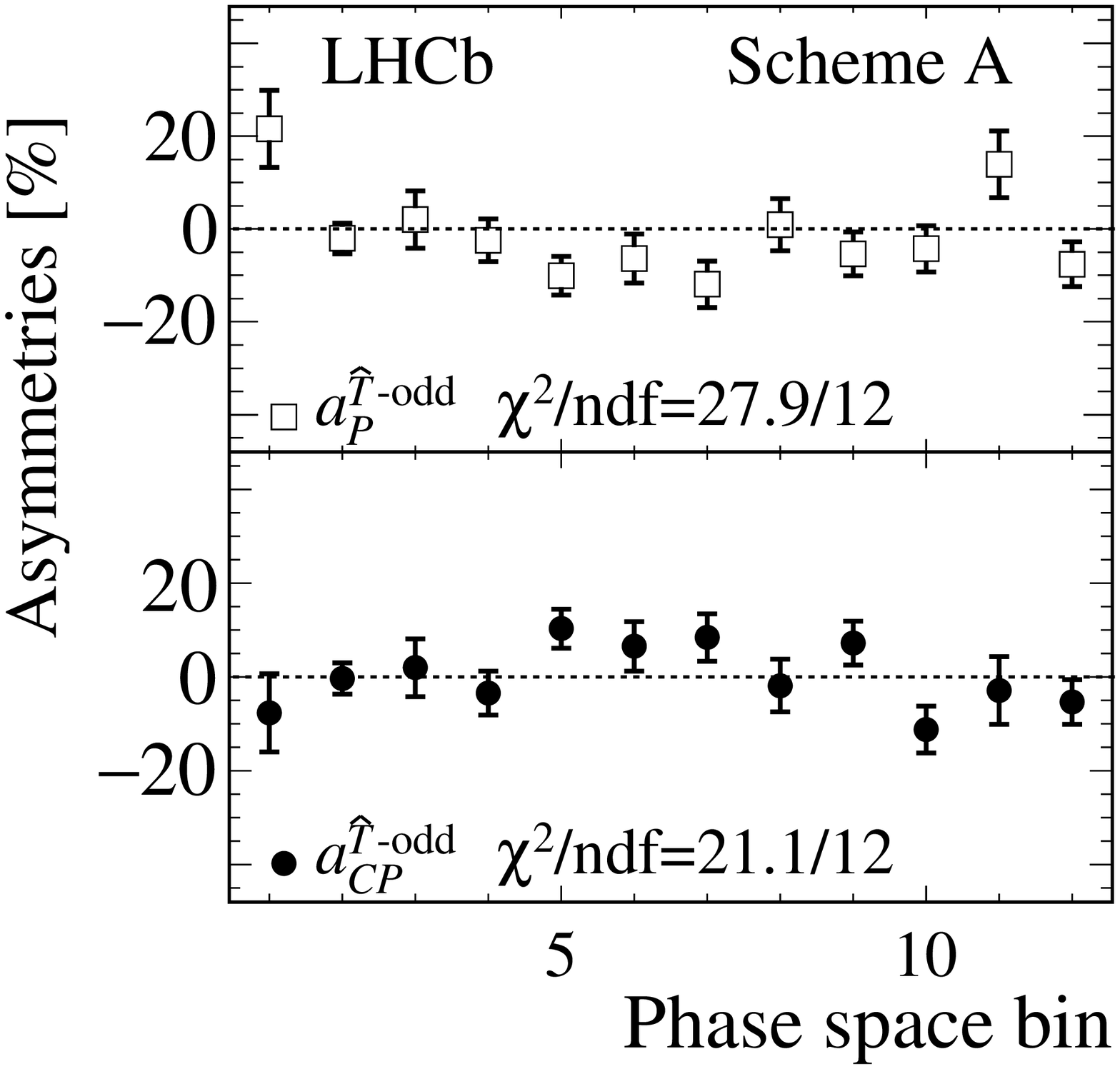}
\includegraphics[width=0.48\textwidth]{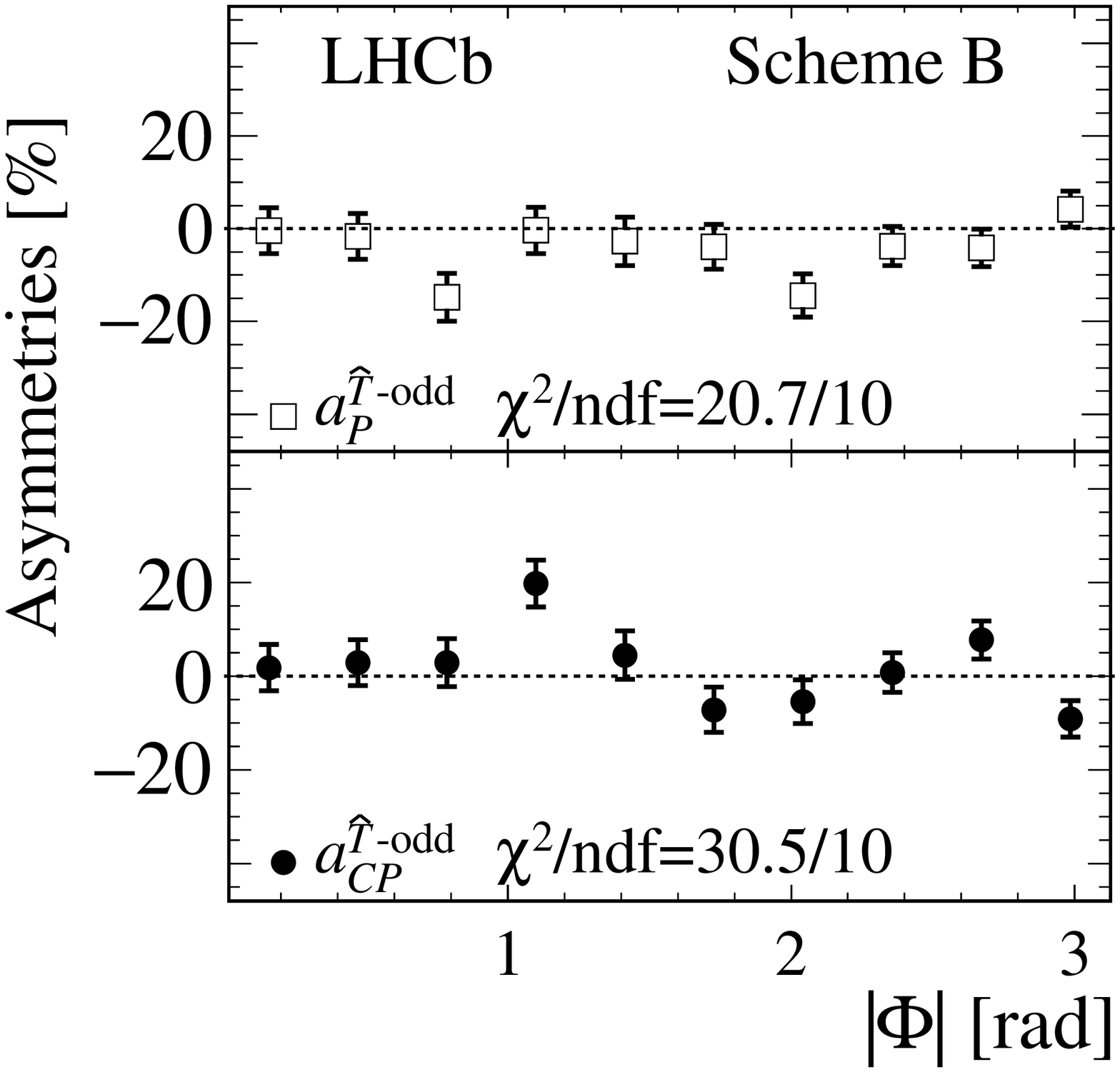}
\caption{\small\label{fig:p3pi_asym} {\bf Distributions of the asymmetries}.
The results of the fit in each region of binning schemes A and B are shown.
The asymmetries \aPTodd and \aCPTodd for \LbToppipipi decays are represented by open boxes and filled circles, respectively.
The error bars indicate the total uncertainties, calculated as the sum in quadrature of the statistical uncertainty resulting from the fit to the invariant mass distribution and the systematic uncertainties estimated as described in the main text.
The values of the \chisq/ndf are quoted for the $P$- and \CP-conserving hypotheses for each binning scheme.}
\end{figure}

The asymmetries measured in \LbToppipipi decays with these two binning schemes are shown in Fig.~\ref{fig:p3pi_asym} and reported in Table~\ref{tab:PHSP_Asym}, together with the integrated measurements.
For each scheme individually, the compatibility with the \CP-symmetry hypothesis is evaluated by means of a \chisq test, with $\chisq=R^T V^{-1}R$, where $R$ is the array of \aCPTodd measurements and $V$ is the covariance matrix, which is the sum of the statistical and systematic covariance matrices.
An average systematic uncertainty, whose evaluation is discussed below, is assigned for all bins.
The systematic uncertainties are assumed to be fully correlated; their contribution is small compared to the statistical uncertainties.
The $p$-values of the \CP-symmetry hypothesis are $4.9\times10^{-2}$ and $7.1\times 10^{-4}$ for schemes A and B,
respectively,  corresponding to statistical significances of $2.0$ and $3.4$ Gaussian standard deviations ($\sigma$).
A similar \chisq test is performed on \aPTodd measurements with $p$-values for the $P$-symmetry hypothesis of $5.8\times10^{-3}$ ($2.8\sigma$) and $2.4\times10^{-2}$ (2.3$\sigma$), for scheme A and B, respectively.
\begin{table}[ht]
\centering
\small
\caption{\label{tab:PHSP_Asym} {\bf Measurements of \CP- and $P$-violating observables.} The \CP- and $P$-violating observables, \aCPTodd and \aPTodd, resulting from the fit to the data are listed with their statistical and systematic uncertainties.
Each value is obtained through an independent fit to a region of the phase space as described in the text and Table~\ref{tab:definitionscheme_Lb2p3pi}.
Results for schemes A and B are outlined for \LbToppipipi decays, and in two bins of phase space for \LbToppiKK decays, as defined in the text. The first column lists the bin number.
For both decay modes the measurement integrated over the phase space, performed independently, is also shown.
}
\input detailedResults_Lb2p3h_tab.tex
\end{table}
The overall significance for \CPV in \LbToppipipi decays from the results of schemes A and B is determined by means of a permutation test~\cite{Stat:Fisher}, taking into account correlations among the results.
A sample of 40,000 pseudoexperiments is generated from the data by assigning each event a random
\Lb/\Lbbar flavour such that \CP symmetry is enforced.
The sign of \CT is unchanged if a \Lb candidate stays \Lb and reversed if the \Lb candidate becomes \Lbbar.
The $p$-value of the \CP-symmetry hypothesis is determined as the fraction of pseudoexperiments with \chisq larger
than that measured in data.
Applying this method to the \chisq values from schemes A and B individually, the $p$-values obtained agree with those from the \chisq test within the uncertainty due to the limited number of pseudoexperiments.
To assess a combined significance from the two schemes, the product of the two $p$-values measured in data is compared with the distribution of the product of the $p$-values of the two binning schemes from the pseudoexperiments. The fraction of pseudoexperiments whose
$p$-value product is smaller than that seen in data determines the overall $p$-value of the combination of the two schemes~\cite{Behnke:2013pga}.
An overall $p$-value of $9.8\times 10^{-4}$ (3.3$\sigma$) is obtained for the \CP-symmetry hypothesis, including systematic uncertainties.

For the \LbToppiKK decays, the smaller purity and signal yield of the sample do not permit $\PV$ and \CPV to be probed with the same precision as for \LbToppipipi, and therefore only two regions of phase space are considered.
One spans $1.43<m(\proton\Km)<2.00\gevcc$ (bin 1) and is dominated by excited \Lz resonances decaying to $\proton\kaon$ and the other  covers the remaining phase space, $2.00<m(\proton\Km)<4.99\gevcc$ (bin 2).
The observables measured in these regions are given in Table~\ref{tab:PHSP_Asym} and are consistent with \CP and $P$ symmetry.

The main sources of systematic uncertainties for both \proton\pim\pip\pim and \proton\pim\Kp\Km decays are experimental
effects that could introduce biases in the measured asymmetries.
This is tested by measuring the asymmetry \aCPTodd, integrated over phase space and in various phase space regions,
using the control sample \LbToLcpiCS, which is expected to exhibit negligible \CPV.
The results are in agreement with the \CP-symmetry hypothesis;
an uncertainty of 0.31\% is assigned as a systematic uncertainty for the \aCPTodd and \aPTodd integrated measurements;
an uncertainty of 0.60\%, the largest asymmetry from a fit to scheme B measurements using a range of efficiency and fit models, is assigned for the corresponding phase space measurements.
The systematic uncertainty arising from the experimental resolution in the measurement of the triple products \CT and \CTbar, which could introduce a migration of events between the bins, is estimated from simulated samples of \LbToppipipi and \LbToppiKK decays where neither $P$- nor \CP-violating effects are present.
The difference between the reconstructed and generated asymmetry is taken as a systematic uncertainty due to this effect and is less than 0.06\% in all cases.
To assess the uncertainty associated with the fit models, alternative functions are used; these tests lead only to small changes in the asymmetries, the largest being 0.05\%.
For \LbToppiKK decays, this contribution is larger, about 0.28\% for the \aCPTodd and \aPTodd asymmetries.

Further cross-checks are made to investigate the stability of the results with respect to different periods
of recording data, different polarities of the spectrometer magnet, the choice made in the selection of multiple candidates, and the effect of the trigger and selection criteria.
Alternative binning schemes are studied as a cross-check, such as using 8 or 12 bins in $|\Phi|$ for \LbToppipipi decays.
For these alternative binning schemes, the significance of the \CPV measurement of the modified scheme B is reduced to below $3\sigma$.
Nonetheless, the overall significance of the combination of these two additional binnings with schemes A and B remains above three standard deviations, with a $p$-value of $1.8\times 10^{-3}$ ($3.1\sigma$), consistent with the $3.3\sigma$ result seen in the baseline analysis.
An independent analysis of the data based on alternative selection criteria
confirmed the results.
It used a similar number of events, of which 73.4\% are in common with the baseline analysis, and gave $p$-values
for \CP symmetry of $3.4\times 10^{-3}$ ($2.9\sigma$) for scheme A and $1.4\times 10^{-4}$ ($3.8\sigma$) for scheme B.

%% file: binning_definition_tab.tex
\begin{tabular}{ccccc}
\hline
\multirow{2}{*}{Phase space bin} & \multirow{2}{*}{$m(p\pi^+)$} &  \multirow{2}{*}{$m(p\pi^{-}_{\text{slow}})$}  & \multirow{2}{*}{$m(\pi^+\pi^-_{\text{slow}})$, $m(\pi^+\pi^-_{\text{fast}})$} &  \multirow{2}{*}{$|\Phi|$}\\
\\
\hline
1  & $ (1.07 , 1.23) $   & & & $(0, \frac{\pi}{2})$     \\
2  & $ (1.07 , 1.23) $  & & & $(\frac{\pi}{2}, \pi)$   \\
3  & $ (1.23 , 1.35) $ & & &  $(0, \frac{\pi}{2})$    \\
4  & $ (1.23 , 1.35) $ & & & $(\frac{\pi}{2}, \pi)$    \\
5  & $ (1.35 , 5.34)$     & $(1.07, 2.00)$					  & $m(\pi^+\pi^-_{\text{slow}})<0.78\textrm{ or\phantom{d}}m(\pi^+\pi^-_{\text{fast}})<0.78$   & $(0, \frac{\pi}{2})$   \\
6  & $ (1.35 , 5.34)$     & $(1.07, 2.00)$					  & $m(\pi^+\pi^-_{\text{slow}})<0.78 \textrm{ or\phantom{d}} m(\pi^+\pi^-_{\text{fast}})<0.78$   & $(\frac{\pi}{2}, \pi)$   \\
7  & $ (1.35 , 5.34)$     & $(1.07, 2.00)$					  & $m(\pi^+\pi^-_{\text{slow}})>0.78\textrm{ and }m(\pi^+\pi^-_{\text{fast}})>0.78$   & $(0, \frac{\pi}{2})$   \\
8  & $ (1.35 , 5.34)$     & $(1.07, 2.00)$					  & $m(\pi^+\pi^-_{\text{slow}})>0.78\textrm{ and }m(\pi^+\pi^-_{\text{fast}})>0.78$   & $(\frac{\pi}{2}, \pi)$   \\
9  & $ (1.35 , 5.34)$     & $(2.00, 4.00)$					  & $m(\pi^+\pi^-_{\text{slow}})<0.78\textrm{ or\phantom{d}}m(\pi^+\pi^-_{\text{fast}})<0.78$   & $(0, \frac{\pi}{2})$   \\
10  & $ (1.35 , 5.34)$     & $(2.00, 4.00)$					  & $m(\pi^+\pi^-_{\text{slow}})<0.78\textrm{ or\phantom{d}}m(\pi^+\pi^-_{\text{fast}})<0.78$   & $(\frac{\pi}{2}, \pi)$   \\
11  & $ (1.35 , 5.34)$     & $(2.00, 4.00)$					  & $m(\pi^+\pi^-_{\text{slow}})>0.78\textrm{ and }m(\pi^+\pi^-_{\text{fast}})>0.78$   & $(0, \frac{\pi}{2})$   \\
12  & $ (1.35 , 5.34)$     & $(2.00, 4.00)$					  & $m(\pi^+\pi^-_{\text{slow}})>0.78\textrm{ and }m(\pi^+\pi^-_{\text{fast}})>0.78$   & $(\frac{\pi}{2}, \pi)$   \\
\hline
\end{tabular}

%% file: detailedResults_Lb2p3h_tab.tex
\newcommand*{\myalign}[2]{\multicolumn{1}{#1}{#2}}
\begin{tabular}{crr}
  \hline
  & \aPTodd [\%]  & \atv [\%]  \\
  &  & \\
Scheme A    &                                \multicolumn{2}{c}{\LbToppipipi}                          \\
\hline
1  & $21.64\pm 8.28 \pm0.60$     &     $-7.69\pm 8.28 \pm0.60$   \\
2  & $-2.04\pm 3.26 \pm0.60$    &     $-0.33\pm 3.26 \pm0.60$   \\
3  & $2.03\pm 6.12 \pm0.60$     &      $1.94\pm 6.12 \pm0.60$   \\
4  & $-2.45\pm 4.60 \pm0.60$     &    $-3.49\pm 4.60 \pm0.60$   \\
5   & $-10.04\pm 4.13 \pm0.60$     &     $10.29\pm 4.13 \pm0.60$   \\
6   & $-6.40\pm 5.23 \pm0.60$      &     $6.51\pm 5.23 \pm0.60$    \\
7   & $-11.91\pm 5.00 \pm0.60$     &    $8.40\pm 5.00 \pm0.60$    \\
8  & $0.94\pm 5.60 \pm0.60$         &      $-1.88\pm 5.60 \pm0.60$   \\
9   & $-5.38\pm 4.67 \pm0.60$       &     $7.20\pm 4.67 \pm0.60$    \\
10 & $-4.26\pm 4.98 \pm0.60$     &     $-11.24\pm 4.98 \pm0.60$  \\
11  & $13.94\pm 7.19 \pm0.60$      &      $-2.90\pm 7.19 \pm0.60$    \\
12  & $-7.64\pm 4.79 \pm0.60$       &     $-5.35\pm 4.79 \pm0.60$   \\
\hline
Scheme B     &               &                                \\
\hline
 1  & $-0.42\pm 4.92 \pm0.60$    &      $\phantom{-0}1.81\pm 4.92 \pm0.60$  \\
 2  & $-1.63\pm 4.88 \pm0.60$   &        $\phantom{-0}2.86\pm 4.88 \pm0.60$  \\
 3  & $-14.73\pm 5.13 \pm0.60$      &      $\phantom{-}2.87\pm 5.13 \pm0.60$   \\
 4  & $-0.32\pm 4.95 \pm0.60$       &      $\phantom{-}19.79\pm 4.95 \pm0.60$  \\
 5  & $-2.71\pm 5.16 \pm0.60$       &        $\phantom{-}4.47\pm 5.16 \pm0.60$   \\
 6  & $-3.85\pm 4.79 \pm0.60$       &       $-7.23\pm 4.79 \pm0.60$             \\
 7  & $-14.40\pm 4.65 \pm0.60$      &       $-5.44\pm 4.65 \pm0.60$             \\
 8  & $-3.75\pm 4.14 \pm0.60$         &      $\phantom{-}0.76\pm 4.14 \pm0.60$   \\
 9  & $-4.16\pm 4.01 \pm0.60$        &      $\phantom{-}7.74\pm 4.01 \pm0.60$   \\
 10 & $4.21\pm 3.84 \pm0.60$           &      $-9.16\pm 3.84 \pm0.60$             \\
\hline
Integrated  & $-3.71\pm1.45\pm0.32$   &      $1.15\pm1.45\pm0.32$  \\
\hline
 &   &       \\
Phase space bin       &                           \multicolumn{2}{c}{\LbToppiKK}                          \\
\hline
1     & $3.27\pm 6.07 \pm 0.66$    &     $-4.68\pm 6.07 \pm 0.66$\\
2     & $4.43\pm 6.73 \pm 0.66$     &    $\phantom{-}4.73\pm 6.73 \pm 0.66$\\
\hline
Integrated  & $3.62\pm4.54\pm0.42$ & $-0.93\pm4.54\pm0.42$  \\
\hline
\end{tabular}

%% file: conclusions.tex
In conclusion, a search for $P$ and \CP violation in \LbToppipipi and \LbToppiKK decays is performed on signal yields of $6646\pm105$ and $1030\pm56$ events.
This is the first observation of these decay modes.
Measurements of asymmetries in the entire phase space do not show any evidence of $P$ or \CP violation.
Searches for localised $P$ or \CP violation are performed by measuring asymmetries in different regions of the phase space.
The results are consistent with \CP symmetry for \LbToppiKK decays but evidence for \CP violation at the $3.3\sigma$ level is found in \LbToppipipi decays.
No significant $P$ violation is found.
This represents the first evidence of \CP violation in the baryon sector and indicates an asymmetry between baryonic matter and antimatter.

%% file: acknowledgements.tex
\section*{Acknowledgements}
\noindent We express our gratitude to our colleagues in the CERN
accelerator departments for the excellent performance of the LHC. We
thank the technical and administrative staff at the LHCb
institutes. We acknowledge support from CERN and from the national
agencies: CAPES, CNPq, FAPERJ and FINEP (Brazil); NSFC (China);
CNRS/IN2P3 (France); BMBF, DFG and MPG (Germany); INFN (Italy);
FOM and NWO (The Netherlands); MNiSW and NCN (Poland); MEN/IFA (Romania);
MinES and FASO (Russia); MinECo (Spain); SNSF and SER (Switzerland);
NASU (Ukraine); STFC (United Kingdom); NSF (USA).
We acknowledge the computing resources that are provided by CERN, IN2P3 (France), KIT and DESY (Germany), INFN (Italy), SURF (The Netherlands), PIC (Spain), GridPP (United Kingdom), RRCKI and Yandex LLC (Russia), CSCS (Switzerland), IFIN-HH (Romania), CBPF (Brazil), PL-GRID (Poland) and OSC (USA). We are indebted to the communities behind the multiple open
source software packages on which we depend.
Individual groups or members have received support from AvH Foundation (Germany),
EPLANET, Marie Sk\l{}odowska-Curie Actions and ERC (European Union),
Conseil G\'{e}n\'{e}ral de Haute-Savoie, Labex ENIGMASS and OCEVU,
R\'{e}gion Auvergne (France), RFBR and Yandex LLC (Russia), GVA, XuntaGal and GENCAT (Spain), Herchel Smith Fund, The Royal Society, Royal Commission for the Exhibition of 1851 and the Leverhulme Trust (United Kingdom).

%% file: LHCb_Authorship_flat_06-Jul-2016.tex
\centerline{\large\bf LHCb collaboration}
\begin{flushleft}
\small
R.~Aaij$^{40}$,
B.~Adeva$^{39}$,
M.~Adinolfi$^{48}$,
Z.~Ajaltouni$^{5}$,
S.~Akar$^{6}$,
J.~Albrecht$^{10}$,
F.~Alessio$^{40}$,
M.~Alexander$^{53}$,
S.~Ali$^{43}$,
G.~Alkhazov$^{31}$
% ...

{\footnotesize \it
% ...
$ ^{5}$Clermont Universit{\'e}, Universit{\'e} Blaise Pascal, CNRS/IN2P3, LPC, Clermont-Ferrand, France\\
$ ^{6}$CPPM, Aix-Marseille Universit{\'e}, CNRS/IN2P3, Marseille, France\\
% ...
$ ^{10}$Fakult{\"a}t Physik, Technische Universit{\"a}t Dortmund, Dortmund, Germany\\
% ...
$ ^{31}$Petersburg Nuclear Physics Institute (PNPI), Gatchina, Russia\\
% ...
$ ^{39}$Universidad de Santiago de Compostela, Santiago de Compostela, Spain\\
$ ^{40}$European Organization for Nuclear Research (CERN), Geneva, Switzerland\\
% ...
$ ^{43}$Nikhef National Institute for Subatomic Physics, Amsterdam, The Netherlands\\
% ...
$ ^{48}$H.H. Wills Physics Laboratory, University of Bristol, Bristol, United Kingdom\\
% ...
$ ^{53}$School of Physics and Astronomy, University of Glasgow, Glasgow, United Kingdom\\
% ...
}
\end{flushleft}